# Nonlinear graphene metasurfaces with advanced electromagnetic functionalities

Boyuan Jin, Christos Argyropoulos*
Department of Electrical and Computer Engineering, University of Nebraska-Lincoln, Lincoln, NE, USA 68588

## ABSTRACT

The optical nonlinear effects can provide different advanced electromagnetic functionalities, such as wave mixing and phase conjugation, which can be applied in a variety of new applications. However, these effects usually suffer from extremely weak nature and require high input intensity values in order to be excited. Interestingly, the large third order nonlinearity of graphene, along with the strong field confinement stemming from its plasmonic behavior, can be utilized to enhance several relative weak nonlinear effects at infrared (IR) and terahertz (THz) frequencies. Towards this goal, various nonlinear graphene metasurfaces are presented in this work to effectively increase the efficiency of different optical nonlinear effects and, as a result, decrease the required input intensity needed to be excited. In particular, we will show that the efficiency of four-wave mixing (FWM) can be improved by several orders of magnitude by using a nonlinear metasurface composed of patterned graphene ribbons, a dielectric interlayer, and a metallic reflector acting as substrate. We also demonstrate that the self-phase modulation (SPM) nonlinear process can be enhanced by using an alternative graphene nonlinear metasurface, operating as coherent perfect absorber, leading to a pronounced shift in the resonant frequency of the coherent perfect absorption (CPA) effect of this structure as the input intensity of the impinging incident waves is increased. This property will provide a robust mechanism to dynamically tune and switch the CPA process. Furthermore, it will be presented that strong negative reflection and refraction can be achieved by a single graphene monolayer film due to the enhancement of another nonlinear process, known as phase conjugation. This nonlinear process is envisioned to be used in the construction of a perfect imaging device with subwavelength resolution.

## 1. INTRODUCTION

Metasurfaces are artificial planar media made of spatially periodic ultrathin unit cells with subwavelength dimensions [1]. They can exhibit novel electromagnetic properties that do not exist in natural occurring materials and have found a plethora of applications, such as perfect absorbers, polarization controllers, cloaking, and nanoantenna devices [2-5]. The majority of the proposed metasurfaces utilize the plasmonic behavior of their metallic parts and exhibit resonant behavior. The electric field is confined in a very small subwavelength-scale region in these cases, where the local electric field intensity can be dramatically enhanced. The electric field enhancement is particularly beneficial to boost different optical nonlinear effects. Hence, the electromagnetic properties of several envisioned nonlinear metasurfaces can be self-tuned by the incident optical intensity via their optical nonlinear response. This effect provides an efficient mechanism to dynamically tune the usually static metasurface response by using new ultrafast self-induced tuning and adaptive mechanisms based on optical nonlinearities. Besides, enhanced optical nonlinear effects can also be of great importance to light manipulation, power amplification, and frequency conversion [6].

In a relevant context, graphene is an ideal material to build nonlinear plasmonic metasurfaces at far-infrared (IR) and terahertz (THz) frequencies compared to noble metals. This is due to the fact that it has a plasmonic response, an exceptionally strong third order nonlinearity, and much smaller losses at far-IR and THz frequency spectra [7-9]. In addition, graphene has a monolayer two-dimensional (2D) structure, which decreases the light–matter interaction volume to extremely subwavelength-scale regions that, subsequently, relaxes the phase-matching requirements of different optical nonlinear parametric effects. Furthermore, both the linear and nonlinear conductivities of graphene are highly tunable via electrical or chemical doping [10, 11]. The resulted tunability provides an additional degree of freedom in the design of graphene metasurfaces and enables them to be adjustable without altering their dimensions and structure.

*christos.argyropoulos@unl.edu; phone 1 402 472-3710; fax 1 402 472-4732; https://argyropoulos.unl.edu

In this work, we present different nonlinear graphene metasurface designs, in particular focusing on four-wave mixing (FWM) enhancement, tunable coherent perfect absorption (CPA), and negative reflection and refraction applications. We demonstrate that by using graphene micro-ribbons above a dielectric interlayer and a metallic substrate, the FWM efficiency will be greatly enhanced by several orders of magnitude. In addition, the frequency where CPA occurs can be dynamically tuned by increasing the input optical intensities when nonlinearities are introduced in a nanostructured graphene film. Furthermore, negative reflection and refraction can be achieved by exciting a special case of the FWM process along a graphene monolayer, known as phase conjugation. The presented results during this work can be used for frequency conversion, optical switching, and super-resolution imaging applications.

## 2. ENHANCED FOUR-WAVE MIXING

In the case of degenerate FWM, the frequencies of the two incident ($f_1$, $f_2$) and the third generated ($f_3$) waves obey the relation $f_3=2f_1-f_2$, where the subscripts 1 and 2 indicate the input pump and signal waves, respectively, and the subscript 3 represents the generated idler wave. The idler wave arises from the third order nonlinear polarization induced by the FWM process in bulk nonlinear materials, which is given by $P_{NL}(f_3) = 6\varepsilon_0 \chi^{(3)}(f_3; f_1, f_1, -f_2) E_1 E_1 E_2^*$, where $\chi^{(3)}$ is the Kerr nonlinear susceptibility and $E_1, E_2$ are the electric fields induced due to the input pump and signal waves, respectively [6]. Graphene is a one-atom thick ultrathin material and is more accurate to be modeled as a surface current instead of a bulk medium [10, 11]. Hence, the graphene's FWM nonlinear contribution can be characterized by a nonlinear surface current given by $J_{NL}(f_3) = 3\sigma^{(3)}(f_3; f_1, f_1, -f_2) E_1 E_1 E_2^*$. The nonlinear surface conductivity of graphene can be approximated to be $\sigma^{(3)} = 3iq^4 v_F^2 / (32\omega^3 \hbar^2 \mu_c)$, where $q$ is the electron charge, $v_F = 1\times10^6$ m/s is the Fermi velocity, $\mu_c$ is the Fermi energy or doping level of graphene, $\omega$ is the radial frequency, and $\hbar$ is the reduced Planck constant [12, 13]. Below the interband transition threshold, which is located at the currently under study low THz frequency spectrum, the linear conductivity of graphene follows a Drude dispersion and can be calculated as $\sigma_L \approx -iq^2 \mu_c / \pi \hbar^2 (\omega - i/\tau)$, where $\tau = \mu\mu_c / qv_F^2$ is the electron relaxation time, and $\mu$ is the DC mobility of graphene [14, 15]. The linear and nonlinear responses of every graphene metasurface in this work are computed based on the aforementioned material properties by using COMSOL Multiphysics, a commercial full-wave electromagnetic solver based on the finite element method.

We start our discussion with a very basic configuration, where a uniform (not patterned) graphene monolayer is placed on top of a 6.3 μm thick dielectric spacer layer and a semi-infinite gold substrate. The dielectric in the interlayer is assumed to have typical values of nonlinear dielectric materials at THz frequencies, which are $\varepsilon_{L,Di} = 4.41$ and $\chi^{(3)}_{Di} = 9.96\times10^{-20}$ m$^2$/V$^2$ [16]. The linear permittivity of gold can be obtained by the Drude model, while the nonlinear response of gold is very weak at THz frequencies and, thus, is neglected [17, 18]. Figure 1(a) shows the distribution of the FWM generated idler power outflow with respect to the incident angles of both the pump and signal input waves, which have frequencies $f_1$=64 THz and $f_2$=62 THz, respectively. The pump and signal waves are both transverse magnetic (TM) polarized and have the same low input intensities $I_1 = I_2 = 30$ kW/cm$^2$. The FWM generated idler wave has frequency $f_3$=66 THz.

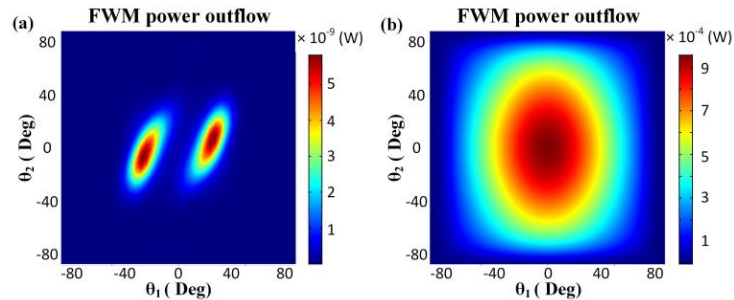

Figure 1. The computed power outflow of the FWM generated idler wave for a uniform graphene film as a function of the incident angles of both pump and signal input waves with parameters: (a) $f_1$=64 THz, $f_2$=62 THz, $\mu_c$=0.1 eV, $\tau = 0.05$ ps and (b) $f_1$=3 THz, $f_2$=2.9 THz, $\mu_c$ =0.3 eV, $\tau = 0.5$ ps.

It can be observed in Fig. 1(a) that the generated idler power outflow is much stronger for some particular incident angles, implying the excitation of surface plasmons along the graphene monolayer surface. For lower input frequencies ($f_1$=3 THz, $f_2$=2.9 THz), the surface plasmons are more difficult to be excited but the nonlinear conductivity of graphene is stronger. Without the presence of surface plasmons, as shown in Fig. 1(b), the FWM generated idler power outflow with frequency $f_3$=3.1 THz has a peak value at normal incidence for both pump and signal input waves. However, the FWM generated idler power values presented in Fig. 1(b) are stronger compared to Fig. 1(a) due to the higher nonlinear conductivity values of graphene at the low THz frequency range.

In order to further increase the FWM efficiency, the uniform graphene layer is patterned to periodic graphene micro-ribbons leading to the creation of a graphene metasurface. As shown in Fig. 2, an array of graphene micro-ribbons are placed over a thin dielectric interlayer and a semi-infinite gold substrate. The inset presents the computed field distribution at the resonant pump frequency of 3 THz, revealing the excitation of highly localized surface plasmon polaritons (SPP) along the graphene ribbons. In addition, the electric field can be further enhanced by Fabry–Pérot (FP) resonances arising from the strong reflection from the bottom metallic (gold) substrate. The width and period of the graphene ribbons are chosen to be 2.91 μm and 3.88 μm, respectively, in order to optimize the maximum absorption to be located at the pump's frequency $f_1$=3 THz. The thickness of the dielectric layer is also chosen to be 6.3 μm. The metasurface can be regarded as uniform along the $z$-axis, because the length of the graphene ribbons is much longer than their width.

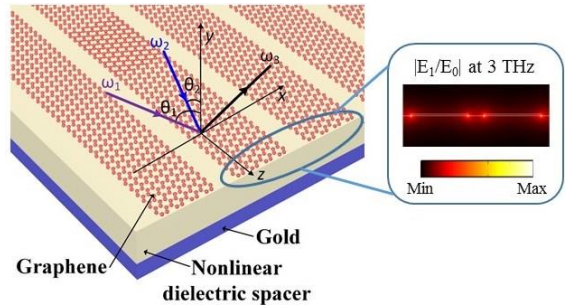

Figure 2. Schematic of the nonlinear graphene metasurface used to enhance FWM. Inset: The normalized electric field enhancement distribution at the resonance frequency of 3 THz, where $E_0$ is the amplitude of the incident electric field.

Under the same illumination conditions and graphene parameters as those used before in Fig. 1(b) ($f_1$=3 THz, $f_2$=2.9 THz, $\mu_c$ =0.3 eV, $\tau$ = 0.5 ps), the power outflow of the FWM generated idler wave is computed and plotted in Fig. 3. The pump and signal waves frequencies are both close to the resonance frequency (3 THz) of the metasurface and their intensities are significantly magnified at the edges of the graphene ribbons, as shown in the inset of Fig. 2(b). As a result, the peak FWM conversion efficiency is enhanced by more than 5 orders of magnitude compared to the uniform graphene monolayer case presented with result before in Fig. 1(b). It can also be seen that the idler power outflow decreases as the angle of incidence of the input waves is increased because less power is coupled into the nonlinear metasurface and more power is reflected back to the surrounding space. However, the idler radiation can still maintain relative large power values within a wide range of incident angles. Note that the structure of the proposed nonlinear graphene metasurface presented in Fig. 2 is simple and easy to fabricate. It can function as a compact and efficient frequency convertor at THz frequencies.

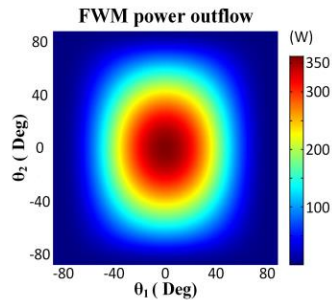

Figure 3. FWM generated power outflow as a function of the incident angles of both pump and signal input waves for the proposed graphene micro-ribbons nonlinear metasurface.

## 3. TUNABLE COHERENT PERFECT ABSORPTION

The coherent perfect absorption (CPA) is a new technique to perfectly absorb light which promises to play a crucial role in new photodetectors design and stealth technology [19-21]. The conventional linear CPA process can be obtained when a lossy ultrathin structure is illuminated by two counter-propagating coherent dual beams [20-25]. The two beams can be generated with a conventional optical interferometer. In principle, the ultrathin slab needs to be slightly lossy in order to obtain interferometric controlled perfect absorption. Hence, perfect absorption is caused by both small material losses and destructive interference between reflected and transmitted waves under bilateral excitations. This process is the time-reversed version of coherent emission (lasing), and CPA is also called anti-laser [24]. The CPA process is envisioned to improve several THz, infrared, and visible nanophotonic components, such as optical switches, modulators, filters, sensors, and thermal emitters. Moreover, the technologies of heating, photovoltaics, water photocatalysis, and artificial photosynthesis depend on the perfect absorption of light; and novel approaches, such as CPA, promise total dissipation of incident electromagnetic energy.

Until now, CPA was obtained with static configurations, and the main limitation was its narrowband response and low reconfigurability. In this work, we design and present a new tunable nonlinear CPA structure based on nonlinear graphene metasurfaces. No outside bias is needed to achieve tunability with the proposed nonlinear devices, and the reconfigurable response is self-induced by the incident electromagnetic radiation due to the metasurface's enhanced nonlinear properties. Figure 4(a) demonstrates a periodically hole patterned graphene film, which can operate as a coherent perfect absorber in its linear operation [21]. It is composed of square holes periodically patterned on a graphene monolayer. The dimensions of the unit cell are shown in Fig. 4(b). The graphene nanoholes period is 400 nm. The holes have a square shape with a side length of 220 nm. The graphene is highly doped and its doping level is $\mu_c$=0.6 eV, with mobility equal to $\mu$=1 m$^2$/V/s leading to an electron relaxation time $\tau = 0.33$ ps. Two counter-propagating incident waves with equal amplitudes input intensities, $I_{i1}$ and $I_{i2}$, respectively, impinge on the graphene metasurface from the left and right sides, as can be seen in Fig. 4(a) (red arrows). The output waves [blue arrows/Fig. 4(a)] exit from the left and right sides of the metasurface and have amplitudes $I_{o1}$ and $I_{o2}$, respectively. The proposed ultrathin metasurface will exhibit CPA in its linear operation when the phase difference of the incident waves, $I_{i1}$ and $I_{i2}$, is either 0 or 180 degrees [22-25].

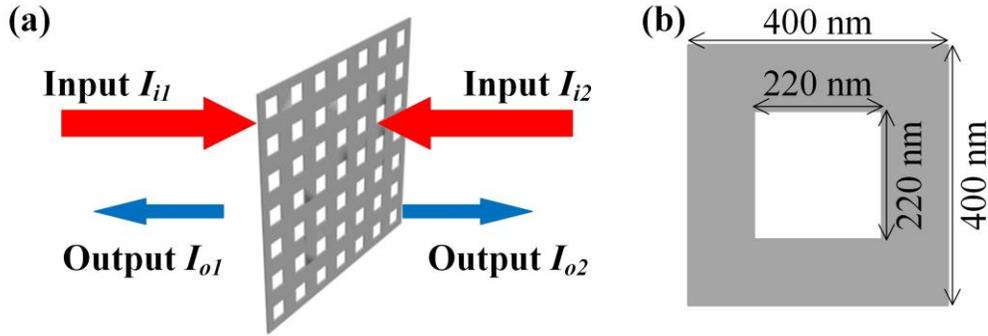

Figure 4. (a) Schematic of the tunable coherent perfect absorber based on a periodically patterned nonlinear graphene film. (b) The unit cell dimensions of the proposed nonlinear metasurface made of periodic square holes patterned on graphene.

When we increase the intensities of both input waves $I_{i1}$ and $I_{i2}$, the nonlinear process of self-phase modulation (SPM) is triggered leading to tunable CPA, with results shown in Fig. 5. This nonlinear process generates a nonlinear surface current $J_{NL} = \sigma^{(3)} |E|^2 E$ along the graphene metasurface, which results in a strong change in the graphene conductivity. The presented nonlinear effect provides an efficient self-induced mechanism to dynamically tune and switch CPA as we vary the input intensity of both incident waves. The frequency response of the proposed coherent perfect absorber is shown in Fig. 5(a). The vertical axis represents the output coefficient, which is calculated by computing the ratio between the total (output) radiated power and the total input power. Zero output coefficient means CPA while high output coefficient values, close to one, mean perfect transmission, i.e., the exact opposite of CPA. It can be seen that for small input intensities $I_{i1} = I_{i2} = I = 1$ W/cm$^2$ the nonlinear effect is too weak to be observed and CPA occurs at 35.41 THz that coincides with the linear plasmonic resonance of the patterned graphene film. The SPM effect becomes stronger as we increase the input intensities and the frequency corresponding to the absorption dip is red-shifted

[Fig. 5(a)]. When the input intensity reaches a low threshold value $I = 10$ kW/cm$^2$, the CPA absorption dip shifts up to 0.16 THz. If we monitor the output powers at a particular fixed frequency, the shift of the spectrum can also lead to a strong variation in the output coefficient values and, as a result, to the CPA effect. Figure 5(b) shows how the output coefficient varies with the input intensity at a fixed frequency of 35.41 THz. The output coefficient is very small (CPA) for low input intensities, when the input frequency coincides with the plasmonic resonance of the graphene metasurface. By increasing the input intensity, especially when exceeding $I = 400$ W/cm$^2$, the frequency shift of the absorption dip becomes significant and the output power at 35.41 THz rapidly rises. Hence, the CPA effect ceases to exist and almost perfect transmission is obtained from both sides of the proposed nonlinear metasurface. Such an efficient control of light with light is unprecedented and can lead to extremely subwavelength all-optical switches operating at THz frequencies.

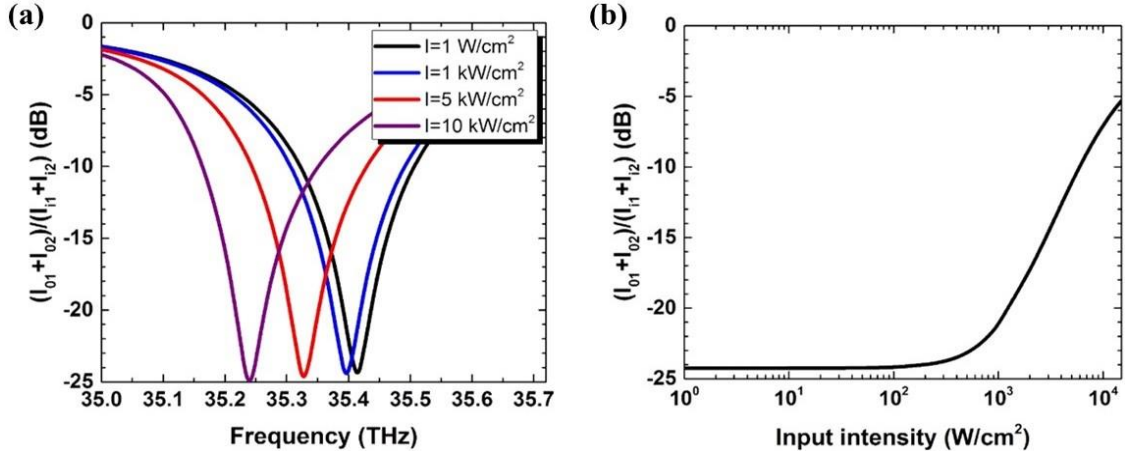

Figure 5. (a) Tunable CPA based on self-phase modulation of the proposed nonlinear graphene metasurface achieved with small input intensities on the order of a few kW/cm$^2$. (b) The dynamic change from CPA to complete transmission when the input intensity is increased and the frequency of operation is fixed to $f$=35.41 THz.

## 4. NEGATIVE REFLECTION AND REFRACTION

Diffraction sets the lower limit on the resolution of optical imaging systems. For any object smaller than half wavelength, its imaging information will be stored in the evanescent waves, which cannot be restored by conventional lenses. It has been demonstrated that a planar negative-index slab can overcome this problem and achieve unlimited resolution [26-28]. Recently, it was also shown that negative refraction can be achieved by using the nonlinear process of phase-conjugation, i.e., a special case of the FWM nonlinear process [29, 30]. Interestingly, graphene is a 2D material with sub-nanometer thickness. Due to this fascinating property, the phase-matching constraints imposed by nonlinear parametric processes, such as FWM, will no longer be relevant to the vertical components of the wave vectors but will only depend on the transverse components along graphene [31]. Therefore, the negative refraction feature can be realized with uniform nonlinear graphene films, due to their strong nonlinear properties, by using nonlinear parametric processes and, in particular, phase conjugation [32]. The proposed configuration to achieve this interesting effect is schematically illustrated in Fig. 6, where the two counter-propagating pump waves (blue arrows) impinge at normal incidence and from opposite directions to the graphene film. In this case, if a probe wave (purple arrow) is obliquely injected onto the graphene surface and all the input waves have the same frequency $f$, the transverse components of their wave vectors will fulfill the relation $k_\parallel^{in} = -k_\parallel^{out}$ according to the phase-matching requirements of this special FWM process (phase conjugation). In this case, the pump and probe waves will be superimposed due to FWM leading to the creation of the output reflected phase conjugated and negative refracted signals (green arrows). Therefore, two new waves will be generated with the same frequency $f$, where one propagates along the probe, but in the backward direction, and the other propagates as if the probe travels along a medium with negative refractive index $n = -1$. The two generated waves will demonstrate negative reflection and refraction features.

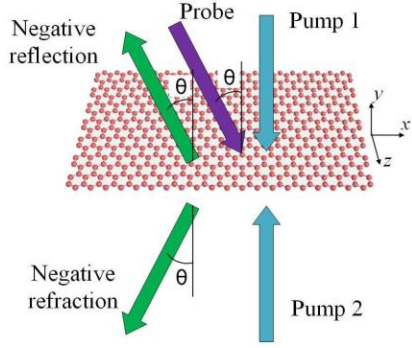

Figure 6. Schematic of the negative reflection and refraction mechanism realized by phase conjugation in a nonlinear graphene film.

Next, we simulate the phase conjugation process in COMSOL Multiphysics, when the incident probe wave is a Gaussian beam, to clearly demonstrate the negative refraction effect. The graphene sample is designed to have the following parameters: $\mu_c$=0.3 eV and $\tau = 0.5$ ps. All inputs are TM polarized waves and have frequencies equal to 150 THz, while the incident angle of the input probe wave is 30°. The effect is simulated by incorporating the nonlinear FWM surface current, given previously in section 2, induced along the graphene monolayer. Figures 7(a) and (b) depict the field distribution and far-field radiation pattern, respectively, of the generated reflected and refracted waves for this nonlinear procedure. The resulted reflection and transmission angles are both 30°, while the incident probe wave along with the reflected and refracted (transmitted) waves are all at the left side of a surface normal to the nonlinear graphene monolayer. Hence, negative reflection and refraction is obtained due to the proposed phase conjugation effect, surprisingly achieved by an extremely thin graphene monolayer. This response is ideal to produce subwavelength imaging, beyond the limits of conventional optical devices.

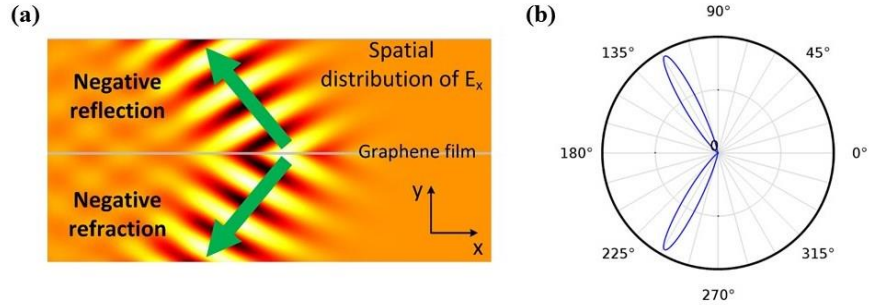

Figure 7. (a) The electric field ($E_x$) distribution and (b) the far-field radiation pattern of the reflected and refracted waves when a Gaussian beam impinges on the nonlinear graphene film. The incident angle of the probe wave is 30° in the presented phase conjugation process.

Based on the aforementioned negative refraction scheme, a subwavelength imaging system is proposed which is composed of two graphene layers, as depicted in Fig. 8. The two incident pumps are both plane waves that have the same intensity. The light departing from the dipole source [dot in Fig. 8] at $x = -d/2$ distance experiences negative refraction at each graphene layer, due to the phase conjugation process, and refocuses to two image points at $x = d/2$ and $x = 3d/2$ distances, where $d$ is the arbitrary distance between the two graphene monolayers. The proposed nonlinear two-graphene layer design can decrease image distortion and increase resolution [27]. We currently work to demonstrate the subwavelength resolution of the proposed nonlinear imaging system and the results will be presented in our future work. Our goal is to demonstrate that it is possible to achieve practically flat unitary transmission and focusing with subwavelength resolution, by using the device presented in Fig. 8, for an arbitrary object illuminated at THz frequencies, resolving details much smaller than the wavelength, contained in the evanescent portion of the angular spectrum of the object. The ultrathin features of graphene ensures the possibility of picking up signals extremely close to the object to be imaged, further enhancing the resolution limit well beyond the limit of classical lenses, even for large distance $d$ between the graphene monolayers. These findings may be groundbreaking for imaging applications; and they are currently being explored in our group with the goal of bringing THz sub-resolution imaging closer to realization.

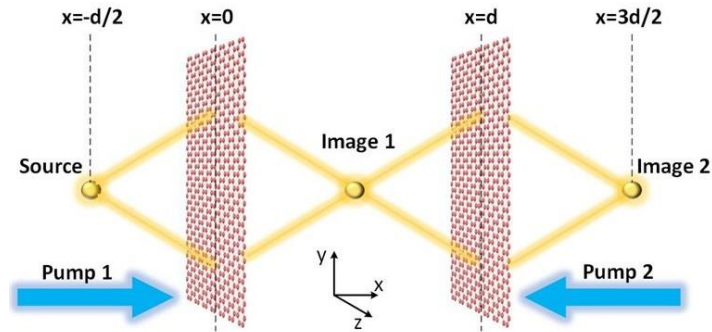

Figure 8. The envisioned setup to achieve subwavelength imaging based on phase conjugation by combining two graphene monolayers with strong nonlinear properties.

## 5. CONCLUSIONS

Graphene is an important material in building nonlinear metasurfaces at infrared and THz frequencies due to its large nonlinear susceptibility, small optical loss, and tunable conductivity by electrical or chemical doping. In addition, the plasmonic behavior of graphene can tightly confine the electric field in small volumes, increase the local electric field intensity, and thus significantly enhancing the efficiency of various optical nonlinear effects. In this work, different nonlinear graphene metasurfaces are presented with the goal to achieve enhanced FWM, tunable CPA, and negative refraction. In particular, it was demonstrated that the FWM efficiency can be increased by several orders of magnitude by using a metasurface composed of graphene micro-ribbons. In addition, the output coefficient, as well as, the absorption spectrum was tuned by varying the input intensity impinging on a nonlinear graphene CPA metasurface due to the SPM nonlinear effect. Finally, negative reflection and refraction was realized when the phase conjugation process was utilized along a nonlinear graphene film. This nonlinear process can lead to the construction of a novel ultrathin imaging system that can achieve subwavelength resolution. The results of this work underlie the wide potential applications of the proposed nonlinear graphene metasurfaces, which can be used to frequency mixing, optical switching, and super-resolution imaging to name a few.

## ACKNOWLEDGMENTS

This work was partially supported by the Office of Research and Economic Development at University of Nebraska-Lincoln, the National Science Foundation (NSF) through the Nebraska Materials Research Science and Engineering Center (MRSEC) (grant No. DMR-1420645), and the Nebraska's Experimental Program to Stimulate Competitive Research (EPSCoR).